\documentclass[12pt]{iopart}
\usepackage{iopams}
\usepackage{graphicx}
\usepackage{xspace,xcolor}
\usepackage{bm}
\usepackage{siunitx}
\usepackage{booktabs}
\usepackage[backend=biber,style=chem-angew,sorting=none,maxnames=2,minnames=1,doi=true]{biblatex}
\usepackage{orcidlink}
\DeclareBibliographyAlias{unpublished}{article}
\DeclareBibliographyAlias{TechReport}{article}

\usepackage{etoolbox}
\makeatletter
\newcommand{\mainmatter}{%
  \setcounter{footnote}{0}%
  \patchcmd{\@makefntext}{\fnsymbol}{\arabic}{}{}%
  \patchcmd{\@thefnmark}{\fnsymbol}{\arabic}{}{}%
  \def\@makefnmark{\textsuperscript{\arabic{footnote}}}%
}
\makeatother

\newcolumntype{C}{S[table-align-text-post=false, table-number-alignment=center]}

\newcommand{\MSun}{M_{\odot}}

\newcommand{\grhayl}{\texttt{GRHayL}\xspace}
\newcommand{\baikal}{\texttt{Baikal}\xspace}

\newcommand{\nrpy}{\texttt{NRPy}\xspace}
\newcommand{\etk}{\texttt{Einstein Toolkit}\xspace}

\newcommand{\mesa}{\texttt{MESA}\xspace}
\newcommand{\rnsid}{\texttt{RNSID}\xspace}
\newcommand{\enzo}{\texttt{Enzo}\xspace}
\newcommand{\rensims}{\texttt{RenSims}\xspace}

\newcommand{\rhob}{\rho}

\begin{document}

% FIX IOPART FOOTNOTE MADNESS
\mainmatter

\title{Gravitational-Wave Signatures of Massive Black Hole Formation}

\author{Bernard J.~Kelly \orcidlink{0000-0002-3326-4454}}
\address{Center for Space Sciences and Technology, University of Maryland Baltimore County, Baltimore, MD 21250, USA}
\address{Gravitational Astrophysics Laboratory, NASA Goddard Space Flight Center, Greenbelt, MD 20771, USA}
\ead{bernardk@umbc.edu}

\author{Sarah Gossan \orcidlink{0000-0002-8138-9198}}
\address{Department of Physics and Astronomy, Hofstra University, Hempstead, NY 11549, USA}

\author{Leonardo R.~Werneck \orcidlink{0000-0002-4541-8553}}
\address{Department of Physics, University of Idaho, Moscow, ID 83843, USA}

\author{John Wise \orcidlink{0000-0003-1173-8847}}
\address{Center for Relativistic Astrophysics, School of Physics, Georgia Institute of Technology, Atlanta, GA 30332, USA}

\author{Zachariah B.~Etienne \orcidlink{0000-0003-1173-8847}}
\address{Department of Physics, University of Idaho, Moscow, ID 83843, USA}

\author{Thiago Assump\c{c}\~{a}o \orcidlink{0000-0002-3419-892X}}
\address{Center for Gravitation, Cosmology and Astrophysics, Department of Physics,
University of Wisconsin-Milwaukee, Milwaukee, WI 53211, USA}

\author{Aláine Lee \orcidlink{0000-0003-1173-8847}}
\address{University of Hawai`i at M\={a}noa, Honolulu, HI 96822, USA}

\author{John G. Baker \orcidlink{0000-0002-1462-1233}}
\address{Universit\'{e} de Toulouse, CNRS/IN2P3, L2IT, Toulouse, France}

\vspace{10pt}

\begin{abstract}
Direct-collapse black holes (DCBHs) are an important component of the massive black hole population of the early universe, and their formation and early mergers will be prominent in the data stream of the Laser Interferometer Space Antenna (LISA). However, the population and binary properties of these early black holes are poorly understood, with masses, mass ratios, spins, and orbital eccentricities strongly dependent on the details of their formation, and the properties of the remaining exterior material (baryonic and non-baryonic), which may be substantial to the point of merger.

We report on initial work to simulate the formation, collapse, and/or merger of such DCBH regions in order to extract the resulting gravitational-wave signals.
\end{abstract}


\section{Introduction}

The direct detection of gravitational waves (GWs) from the merger of a stellar-mass black-hole binary by the LIGO observatories in 2015 ushered in a new era of gravitational-wave astronomy. Since then, the LIGO-Virgo-KAGRA (LVK) collaboration has detected more than three hundred compact object mergers; while there have been a couple of confirmed merging neutron-star binaries (GW170817, GW190425) \cite{TheLIGOScientific:2017qsa,LIGOScientific:2020aai}, and black-hole-neutron-star binaries (GW230518\_125908 and GW230529\_181500), the bulk of these detections have been of black-hole binaries with total system masses of ${\sim} 10^1\text{--}10^2 \MSun$, \cite{LIGOScientific:2018mvr,LIGOScientific:2020ibl,KAGRA:2021vkt,LIGOScientific:2025slb}.

LVK and other ground-based interferometric facilities are most sensitive to GWs with frequencies ${\sim}10^1\text{--}\qty[parse-numbers=false]{10^{3}}{\hertz}$, produced in the final orbits and merger of stellar-scale compact bodies.
The largest-mass systems observed to date by the LVK range from ${\sim}140 \MSun$ (GW231028\_153006, GW190521) to ${\sim}230 \MSun$ (GW231123) \cite{LIGOScientific:2021usb,LIGOScientific:2025rsn}.
As GW frequencies scale inversely with the total system mass $M$, such events for compact binaries with masses much higher than this generally fall outside the LVK band, and can only be observed by larger-baseline instruments.

The Laser Interferometer Space Antenna (LISA) \cite{Colpi:2024xhw}, planned to launch in the mid-2030s, is an instrument designed to be sensitive to GW sources with frequencies \mbox{${\sim}0.001$--\qty{1}{\Hz}}, well below those accessible to LIGO and other ground-based facilities.
LISA’s standard sources include merging massive (${\sim}10^5\text{--}10^6 \MSun$) comparable-mass black-hole binaries (MBHBs); stellar-scale objects merging with massive black holes (extreme-mass-ratio inspirals, or EMRIs); and compact stellar-scale binaries (e.g. white dwarf binaries) long before merger.
LISA will also likely encounter a stochastic background of sources of both cosmological and astrophysical origin; the latter includes unresolved stellar-mass black-hole binaries and white-dwarf binaries \cite{Colpi:2024xhw}, as well as possible prompt collapse events \cite{Pacucci:2015tpa}.

The strongest classic LISA sources are all persistent binary sources, with long observable lifetimes, and consequently high accumulated signal-to-noise ratio (SNR).
However, other astrophysical events can generate GW signals in the LISA sensitivity band, albeit shorter-duration signals -- for instance the direct collapse of stellar, proto-stellar, and prestellar material to form black holes.
The resulting direct-collapse black holes (DCBHs) are possible LISA sources both in their formation stage, and as progenitors of the massive black-hole binaries that will later go on to provide merger signals \cite{Pacucci:2015tpa,Natarajan:2016rii}.
Intermediate-mass black holes (IMBHs) resulting from direct collapse are expected to yield merging binaries that may be observed in the dHz band \cite{Kawamura:2011zz,Gong:2021any,Li:2024rnk,Lee:2025qbu}.
As such, they're an important source for planned instruments with peak sensitivity in that band, such as the space-based Deci-hertz Interferometer Gravitational wave Observatory (DECIGO) \cite{Kawamura:2020pcg}, and the Moon-based Laser Interferometer Lunar Antenna (LILA) \cite{Jani:2025uaz} and Lunar Gravitational-wave Antenna (LGWA) \cite{LGWA:2020mma}, and possibly in lower-frequency instruments, such as TianQin \cite{Wang:2025lsl} and TAIJI \cite{Gong:2021any}.

The production of supermassive black holes (SMBHs) expected at the core of most galaxies is an area of active research. Standard methods of producing higher-mass black holes from stellar-scale seeds include accretion of surrounding material and repeated mergers of smaller black holes.
However these channels appear incapable of generating the observed population of SMBHs in the time available in standard cosmological models \cite{Hayes:2024czx}.
This problem has been intensified with recent observations of unexpectedly mature galaxies at high redshift by the James Webb Space Telescope \cite{Pacucci:2023oci}.
Early production of more massive seeds in the IMBH regime, through direct collapse, would help resolve this discrepancy \cite{Inayoshi:2019fun}, and distinguishing this from other channels is a core part of the science case of LISA and other lower-frequency instruments, such as TianQin \cite{Li:2024rnk} and TAIJI \cite{Shen:2025uya}.
There is general agreement on the importance of suppression of molecular hydrogen in keeping temperatures high enough ($\gtrsim 10^4 \textrm{K}$) to resist fragmentation, and promote direct collapse to black holes with masses $\sim 10^5 \MSun$ \cite{Omukai:2000ic,Oh:2001ex,Bromm:2002hb,Koushiappas:2003zn,Lodato:2006hw,Wise:2007bf,Regan:2008rv,Latif:2013dua,Luo:etal:2020,Lu:2024zwa}, though additional exotic cosmological seeding or accelerating mechanisms have been suggested \cite{Eisenstein:1994nh,deJong:2024rnc,Chiu:2025vng}.

Our current knowledge of the GW signals from DCBH formation is restricted to a handful of analytical and quasi-numerical estimates.
For instance, \cite{Suwa:2007du} uses results of a Newtonian 2D hydrodynamics + neutrinos collapse performed with the \texttt{ZEUS-2D} code to generate quadrupole waveforms. The waveform contributions due to matter and neutrinos are generated separately and added to produce a full waveform. 
\cite{Li:2010zzr} performs a Newtonian calculation of approximate collapse waveforms for intermediate-mass black holes, using an oblate spheroidal matter distribution that undergoes free-fall collapse, and tracking the development of the moment of inertia to get waveforms. The collapse waveform is tied to quasinormal ringdown, so by construction, the waveform is simple, and ends with ringdown.
\cite{Pacucci:2015tpa} estimates likely DCBH gravitational-wave signals at cosmological distances, using ``modern waveforms'' based on the work of \cite{Suwa:2007du} and \cite{Li:2010zzr}. Coupled with population estimates, they predict peak amplitude at around \qty{0.9}{\mHz}.

We outline in this paper our approach to studying the formation of DCBHs, and their likely relevance for LISA science, beginning with a more astrophysically realistic starting point, and incorporating multiple computational approaches at different length- and time-scales.

\section{A Multi-Code Approach to DCBH Formation}

Rather than begin with some random or idealized fluid configuration with no connection to astrophysical expectations, we aim to use a three-stage procedure that spans the full range of scales:
\begin{enumerate}
    \item derive initial conditions from cosmological simulations such as the Renaissance Simulations~\cite{OSheaEtAl_2015};
    \item use a Newtonian protostellar evolution code (\mesa) to follow regions of interest until they enter strong-gravity regime;
    \item finish off with full Numerical Relativity (\etk) when gravity is too strong to treat with Newtonian theory.
\end{enumerate}

\subsection{Cosmological Simulations}
\label{ssec:cosmo_overview}

\begin{figure}
\begin{centering}
\includegraphics[width=0.8\textwidth]{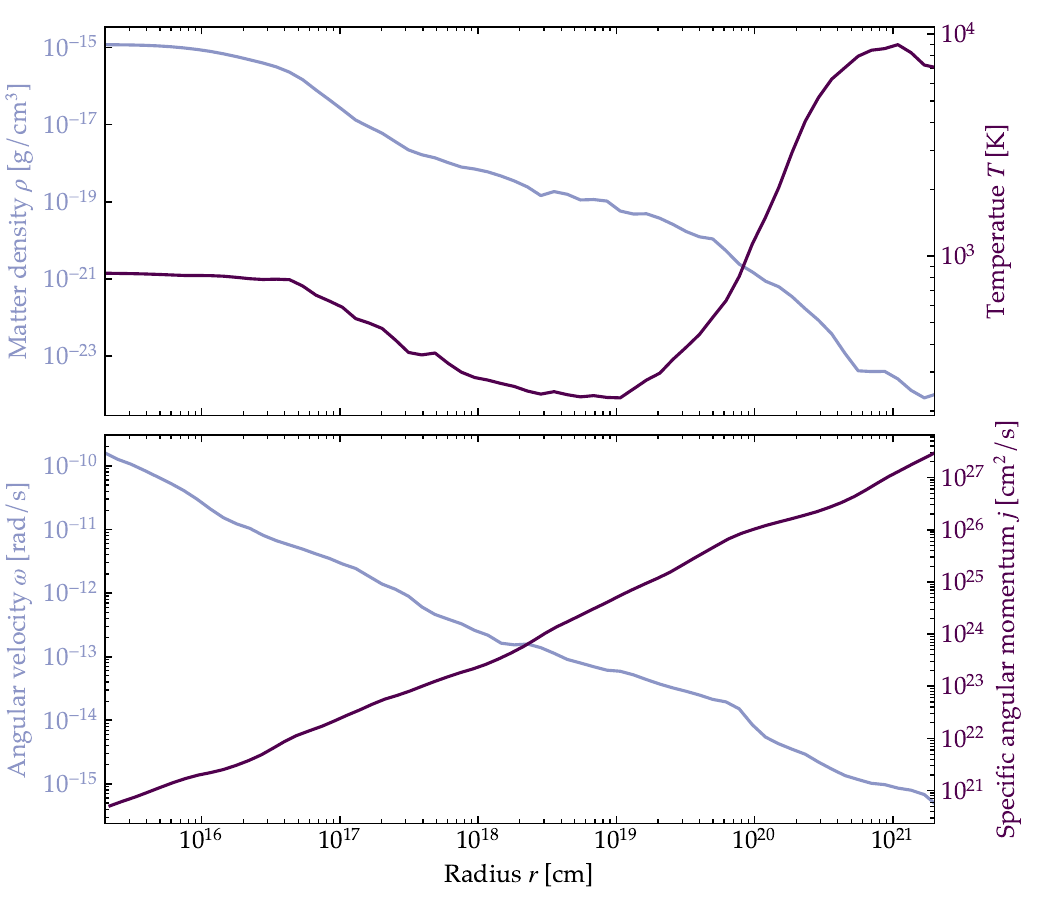}
\caption{\textit{Top}: density and temperature profiles extracted from the LWH clump of \cite{WiseEtAl_2019}. \textit{Bottom}: angular velocity specific angular momentum profiles for the same clump demonstrating a strong differential rotation law.}
\label{fig:RenSims_LWH}
\end{centering}
\end{figure}

The Renaissance Simulations (\rensims) \cite{OSheaEtAl_2015} were performed between 2013--2015 using the \enzo code \cite{ENZO:2013hhu} on the Blue Waters supercomputer at NCSA (National Center for Supercomputing Applications) that simulate the first stars and galaxies within a (40 comoving Mpc)$^3$ cosmological volume at redshifts greater than 10.  It solves the N-body dynamics for dark matter and radiation hydrodynamics equations in a hierarchy of adaptive mesh refinement (AMR) grids that focus on high-density regions of interest. The simulation suite is a rich data set with over 3,000 galaxies and 30,000 primordial stars that can be mined for multiple purposes and research projects.
Our particular interest here is locating regions of gas that are likely to collapse to a single compact object. Several such ``halos'' were identified by \cite{WiseEtAl_2019}, including the ``most massive halo'' and the one exposed to the highest Lyman-Werner radiation flux (LWH).
These halos are metal-free without prior star-formation that exhibit strong radial inflows, conducive to DCBH formation.
From these likely-collapse halos, we extract profiles for density, pressure, and angular momentum, to name a few.
Profiles of density and temperature drawn from the LWH halo are shown in Figure~\ref{fig:RenSims_LWH}. Pressure is not directly supplied from this data, but can be reconstructed assuming ideal-gas behavior; here it would peak at around a few $\qty[parse-numbers=false]{10^{-4}}{dyn/\cm\squared}$.

These \rensims profiles are one-dimensional, showing only radial dependence. Obviously the halos contain rotating fluid, and thus should not be expected to exhibit perfect spherical symmetry. However, the angular velocities are modest at this early stage, and deviations from symmetry are correspondingly small. The largest deviations for the LWH halo occur between $10^{18}\text{--}\qty[parse-numbers=false]{10^{19}}{\cm}$ where there is a flattening in the density profile arising from a filamentary structure forming.
That said, the radial profiles produced are radial averages, and thus efface a lot of fine detail in the halo structure.
The center of each halo is given by the maximum of baryonic density $\rhob$; more details on the halo characterization can be found in \cite{WiseEtAl_2019}.

Note that direct collapse is not guaranteed in previous work identifying collapse regions.
The Renaissance Simulations ended when the baryon density reached $10^{-15} \textrm{g}/\textrm{cm}^3$, and the authors inferred a direct collapse from the end-state of their simulation; in particular the gas infall rates indicated that the collapsing cloud would accrete to a point within a million years.
However this approximation, as they noted, did not include any feedback effects from the resulting collapsed object.

It is tempting to take a likely collapse region from \rensims and feed it directly into a numerical relativity (NR) code. However, there is a large scale discrepancy in this approach. For a typical NR simulation of a black hole (BH) system with total mass $M$, the domain size $L \sim 10^4\text{--}10^5 \, G M/c^2$, and we need to resolve the BH horizon, of size ${\sim} \, G M/c^2$. However, for a \rensims clump of interest with total mass $M \sim 3 \times 10^6 \, \MSun$, the matter is spread over $L \sim \qty{5e2}{pc}$, which is ${\sim}3 \times 10^9 \, G M/c^2$. This means we would need to allow ${\sim}10^4$ increase in dynamic range to track collapse to BH formation, compared to standard vacuum NR evolutions.
This huge additional volume of space could be addressed through the addition of 13--14 levels in a fixed (FMR) or adaptive mesh refinement (AMR) scheme, without requiring unreasonable additional memory; each additional coarser level could twice as far in all directions as the previous level, using half the resolution of the previous level, so the total memory required would be a small multiple of a typical NR BH simulation.
However the overall evolution timescale is related to the physical extent of the matter, while the NR timestep taken is constrained by the highest-resolution refinement grid, covering the smallest domain; by increasing the size of the domain by a factor of $10^4$, we also increase the required number of timesteps to simulate the full collapse by a similar factor, which makes the simulation impractical.

\subsection{Mesoscale: Supermassive Star Simulations}
\label{ssec:meso_overview}

To bridge this scale gap, we plan to proceed along a faster and cheaper Newtonian evolution, using Modules for Experiments in Stellar Astrophysics (\mesa) \cite{Paxton:2010ji,Paxton:2013pj,Paxton2015,Paxton2018,Paxton2019,Jermyn2023}.
\mesa is a one-dimensional Lagrangian code that implicitly solves the equations of stellar structure. Inherently multidimensional hydrodynamic processes such as convective mixing and nuclear burning are treated via diffusion.

Initial data inputs into \mesa will be drawn from the physical properties of different halo regions identified in the \rensims profiles: the baryonic (fluid) density $\rhob$, pressure $P$, and temperature $T$, as well as the angular momentum distribution of the material. Primordial gas composition (hydrogen mass fraction $X_{\mathrm{H}} = 0.7516$ and helium mass fraction $X_{\mathrm{He}} = 0.2484$) is assumed. 

To this we must add assumptions about the likely nuclear reaction chains involved in the protostellar evolution in \mesa. Given the comparatively low temperatures ($T \sim 10^{3}\text{--}\qty[parse-numbers=false]{10^{4}}{\K}$) and pressures of the \rensims halo data, and the fact that the \mesa evolution should end before a main-sequence star has formed, a small number of species should be adequate. Our flagship runs employ a 21 species nuclear reaction network, while we run experiments using larger nuclear reaction networks incorporating additional isotopes relevant at high temperatures ($T \geq 10^{9}\,\mathrm{K}$; up to 256 species) to investigate the impact on the pre-collapse stellar structure.

We employ adaptive grid resolution such that fractional differences in $\log T$, $\log \rho$, and $\log X_{i}$ do not exceed \qty{0.5}{\percent}.

\subsubsection{Limitations of Mesoscale Evolution}

The \mesa code contains a sophisticated treatment of stellar evolution processes not available in our NR codes, including extensible nuclear reaction networks and radiative cooling. However, its treatment of gravity is Newtonian, with some post-Newtonian corrections to the potential. This limits how far its results can be trusted as massive stars evolve to higher-density configurations.  

\subsubsection{Output and Analysis}

\mesa outputs the current state of the simulated star in the form of radial profiles of mass, pressure, density, other thermodynamic fields, and the distribution of angular momentum, as well as the detailed species composition of the star.

\subsection{Strong Gravity: Numerical Relativity}

To simulate the final collapse in the regime where Newtonian gravity is no longer adequate, we use the \etk~\cite{Loffler:2011ay} (specifically the 2023\_05 release, codenamed ``Karl Schwarzschild'') for simultaneous evolution of the stellar material and the underlying spacetime.

The fundamental fields of numerical relativity come from the ``3+1'' (three spatial dimensions and one temporal dimension) Arnowitt-Deser-Misner (ADM) split of the 3D spacetime into a foliation of 3D spatial hypersurfaces (``slices'') with hypersurface normals $n^a$. In this approach, the full spacetime four-metric $g_{ab}$ is decomposed into: the \emph{three-metric} $\gamma_{ij}$ representing the geometry of each slice, the \emph{lapse function} $\alpha$ defining the infinitesimal separation of the slices, and the \emph{shift vector} $\beta^i$ defining how the spatial coordinates change from slice to slice. The four-dimensional line element then is given by
\begin{equation*}
    ds^2 = g_{ab} dx^a dx^b = -\alpha^2 dt^2 + \gamma_{ij} (dx^i+\beta ^i dt)(dx^j+\beta ^j dt).
\end{equation*}
(Here and later on, early-alphabet subscripts and superscripts $\{a, b, c, \dots \}$ represent spacetime indices, ranging from 0 to 3; mid-alphabet ones $\{i, j, k, \dots\}$ represent purely spatial indices, ranging from 1 to 3.)
Given the spatial metric $\gamma_{ij}$ we can also define a corresponding covariant derivative $D^i$, a three-connection $\Gamma^i_{ik}$, and curvature fields such as the three-Riemann tensor $R^i_{jkl}$, the three-Ricci tensor $R_{ij} \equiv R^m_{imj}$, and three-Ricci scalar $R \equiv \gamma^{mn} R_{mn}$.
The \emph{extrinsic curvature} $K_{ij}$ is related to the rate of change of the three-metric with coordinate time $t$:
\begin{equation*}
K_{i j}  = \frac{1}{2\alpha} \left( \beta_{i,j} + \beta_{j,i} - \partial _t \gamma_{i j} - 2 \Gamma^p_{i j} \beta_p \right),
\end{equation*}
with trace $K \equiv \gamma^{ij} K_{ij}$.

The matter and energy content of the configuration is encapsulated in the \emph{stress-energy} tensor $T^{a b}$. For an ideal unmagnetized fluid, this takes the form
\begin{equation}
    T^{ab} = \rhob h u^a u^b + P g^{ab} \label{eq:Tab_def}
\end{equation}
where $g^{ab}$ is the inverse of the four-metric $g_{ab}$, $u^a$ is the fluid four-velocity, $\rhob$ is the fluid's baryonic density, $P$ is its pressure, $h \equiv 1 + \epsilon + P/\rhob$ is the specific enthalpy, and $\epsilon$ is the specific internal energy of the fluid.
For a fluid at rest relative to Eulerian observers the four-velocity coincides with the hypersurface normal: $u^a = n^a$.

On each time slice, the metric fields should satisfy the Hamiltonian and momentum constraint equations (up to a conventional factor of $8 \pi$) \cite{Smarr:1977uf,Baumgarte:1998te}
\begin{eqnarray}
    R - K_{ij} K^{ij} + K^2 &= 2 E, \label{eq:CHam_ADM}\\
          D_j K^j_i - D_i K &= S_i, \label{eq:CMom_ADM}
\end{eqnarray}
where the terms on the right contain the corresponding matter-energy distribution on each time slice, derived from the 3+1 decomposition of the stress-energy tensor $T^{ab}$ \eref{eq:Tab_def}:
\begin{eqnarray*}
    E &\equiv T_{ab} n^a n^b, \\
    S_i &\equiv - T_{ic} n^c - E n_{i}.
\end{eqnarray*}

Spacetime evolution is performed using the Baumgarte-Shapiro-Shibata-Nakamura-Oohara-Kojima (BSSNOK) \cite{Nakamura:1987zz,Shibata:1995we,Baumgarte:1998te} formalism as implemented by the \baikal code module~\cite{Etienne:2024ncu}.
More sophisticated variants of the BSSNOK formalism, such as CCZ4 \cite{Bona:2003fj,Bernuzzi:2009ex,Alic:2011gg,Alic:2013xsa} may offer better control of constraint violations or high-frequency noise; we will consider these in future investigations, though because of the low densities and spacetime curvature of our initial conditions compared to, e.g. core-collapse supernovae simulations, we do not expect this to be crucial for qualitatively correct results.
Matter evolution is treated using Ideal General Relativistic Hydrodynamics, as implemented by \grhayl~\cite{Jacques:2024pxh}.
Both \baikal and \grhayl code modules were developed with the \nrpy framework~\cite{nrpy_github}.

For efficient use of resources, we use a multi-level Carpet numerical mesh for all fields, activating extra levels as the collapse proceeds and the curvature increases in the center of the computational domain.
This increase is tracked by evaluating the minimum of the lapse function $\alpha$. Initially $\alpha \approx 1$ everywhere in the domain, but its minimum drops toward zero as matter concentrates, and we activate extra levels of refinement as $\min(\alpha)$ falls past certain trigger values (0.45, 0.3, 0.15, \dots).

\subsubsection{Limitations of Strong Gravity Evolution}

While \grhayl can handle general equations of state (EOS) in tabulated form, it is simpler to assume a barotropic EOS $P = P(\rhob)$, $T = T(\rhob)$, etc., specifically, a polytrope $P = K \rhob^{\Gamma}$.
We note that the \mesa code has its own more sophisticated EOS, a hybrid of rules over different regimes \cite{TimmesSwesty_2000,JermynEtAl_2021}. This is \emph{not} barotropic, and approximating this with a polytrope introduces an additional source of error. However, the \rnsid procedure (see below) is only defined for a barotropic EOS. Moving away from \rnsid in future developments will allow us to adopt a more faithful representation of the \mesa output.

As they deal purely with the hydrodynamic quantities as classical fields the \etk evolutions cannot capture much of the physics handled by \mesa in the Mesoscale state. In particular, there is no treatment of nuclear reactions, or radiation or neutrino transport.

\subsubsection{Output and Analysis}

The main direct product of the evolution is the same fields evaluated at later times, as well as diagnostics calculated from these fields, including the mass and spin of the post-collapse and gravitational radiation decomposed into angular modes.

To find apparent horizons during the evolution, we employ the module (thorn) \texttt{AHFinderDirect} \cite{Thornburg:2003sf}.
To extract gravitational radiation, we calculate the complex radiative Weyl Scalar $\psi_4$ using the \texttt{WeylScal4} thorn, and decompose this field into spin-weighted $(l,m)$ spherical harmonics on spheres of constant coordinate radius $r$ up to $l = 4$ using the \texttt{Multipole} thorn.

\section{Linking Up the Codes}

How do we translate between very different approximations for matter and spacetime used at different scales
(and codes)?

From the Cosmological to the Mesoscale, we need to identify the size and initial chemical composition of the halo in question. As the \mesa code expects spherically symmetric data, an average radial profile for each field of interest should be sufficient. This includes a rotation profile, since \mesa allows for rotation in its stellar evolutions. Since the halo is metal-free by definition, the initial chemical composition of the region is trivial, though we must allow for several species of nuclear products during the subsequent \mesa evolution. 

Going from the Mesoscale to the Strong-Gravity Regime, we need the hydrodynamical (HD) fields as evolved by \mesa to a more compact state. However, since the strong-gravity evolution involves dynamical general relativity, this HD specification must be accompanied by a specification of the space-time metric at the time of hand-off.
The metric fields must satisfy the Hamiltonian and momentum constraints \eref{eq:CHam_ADM}--\eref{eq:CMom_ADM}.

As noted above, the 1D \mesa code allows for rotating configurations. Nevertheless, this is implemented implicitly through the redefinition of the radial coordinate $r$, and this must be accounted for when translating to an explicitly 3D code like the \etk, which expects an explicit $\theta$ (polar angle) dependence for rotating configurations.

One approach to obtaining constraint-satisfying initial metric data is to use the initial data ansatz developed by Komatsu, Eriguchi, Hachisu (KEH) \cite{Komatsu:1989zz,CookEtAl_1992}, and implemented as the \rnsid code \cite{Stergioulas:1994ea}. This approach assumes the spacetime is represented by a stationary four-metric of the form
\begin{equation}
    ds^2 = - e^{2 \nu} dt^2 + e^{2 \mu} (dr^2 + r^2 d\theta^2) + e^{2 \psi} r^2 \sin^2\theta (d\phi - \omega dt)^2,
\end{equation}
where the ``metric potentials'' $\nu(r,\theta)$, $\mu(r,\theta)$, $\psi(r,\theta)$, and $\omega(r,\theta)$ are determined by Einstein's equations, under the assumption that the fluid obeys a polytropic equation of state: $P = K \rhob^{\Gamma}$ (or alternatively a tabulated barotropic one).\footnote{There are several conventions for the names of the metric potentials; we use those of \cite{Stergioulas:1994ea}, to minimize conflicts with other field quantities.}
Note that the \mesa code employs its own more realistic EOS, a hybrid of rules over different regimes \cite{TimmesSwesty_2000}.
This is not barotropic, much less a polytrope; thus, in mapping \mesa outputs to \rnsid inputs, we can approximate the \mesa HD field behavior to a polytrope, by recording the central density $\rhob_c$ and pressure $P_c$ of the evolved \mesa configuration, using adiabatic (i.e. constant specific entropy $s$) index $\Gamma = \Gamma_1 \equiv (d \ln P/ d \ln \rhob)_{s}$, found to be approximately constant over the highest-density interior regions of the configuration.

Under these conditions, the configuration is determined by three freely specified parameters: the central fluid density $\rhob_c$; the polar-to-equatorial radius ratio $r_p/r_e$, which varies from 1 (nonspinning) to 0 (highly spinning); and the differential rotation parameter $\hat{A}$, where $\hat{A} \ll 1$ indicates extremely differential rotation, and $\hat{A} \gg 1$ indicates uniform rotation.
These parameters can be deduced from bulk properties of the \mesa end-state, rather than using an explicit radial profile.

The \rnsid code solves these efficiently on a compactified 2D numerical grid, with regular spacing in $\mu \equiv \cos\theta$ and $s \equiv r / (r + r_e)$, where $r_e$ is taken as the equatorial radius of the stellar surface.
Beginning with the nonspinning case $r_p/r_e = 1$ --- equivalent to the Tolman-Oppenheimer-Volkov (TOV) solution \cite{Tolman:1939jz,Oppenheimer:1939ne} --- and then progressively adding spin by reducing $r_p/r_e$ until it reaches the desired value.
More details on the iterative procedure employed at each step can be found in \cite{CookEtAl_1992}.
The resulting field values are then interpolated onto the computational domain for evolution.
The specific implementation of \rnsid in the \etk is supplied by the thorn \texttt{Hydro\_RNSID}, originally supplied by the authors of \cite{Stergioulas:1994ea}.

\section{Implementing Final Collapse}

Since the \rnsid code produces stationary configurations, some kind of perturbation is necessary to induce collapse of the initial data.
A simple option, used in \cite{Saijo:2009ub}, is to reduce the polytropic parameter $K$ by a few percent after the initial solve, to reduce the pressure support throughout the star.

We test this procedure for \rnsid data by using an initial stellar configuration studied by \cite{Saijo:2009ub}.
In fact, we use a uniformly rotating version of their Model I; using uniform rotation rather than differential results in a much smaller amount of angular momentum in the system.
Once the \rnsid data was calculated and interpolated to the 3D computational grid, we reduced the polytropic $K$ by \qty{5}{\percent} to induce collapse.

\begin{figure}
\begin{centering}
\includegraphics[width=0.8\textwidth]{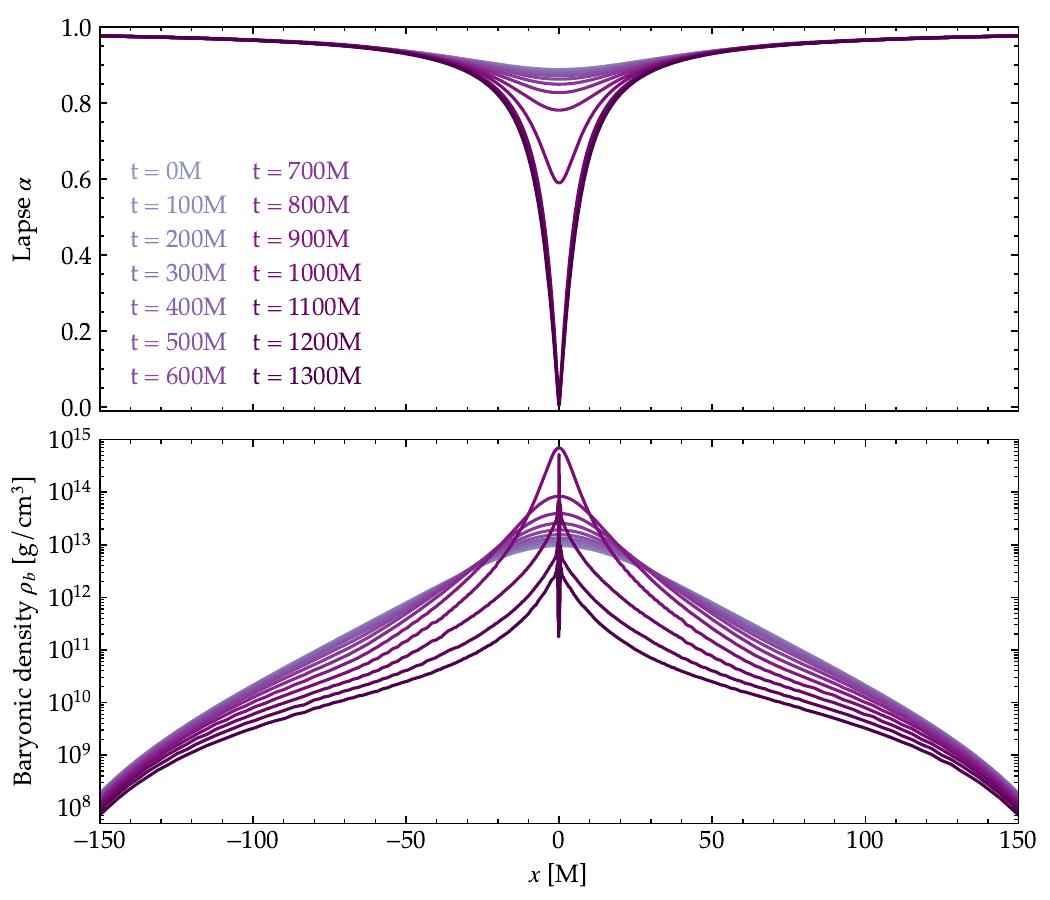}
\caption{\textit{Top}: Lapse function $\alpha$ evolution along $x$-axis from the same simulation. \textit{Bottom}: density evolution along $x$-axis from Saijo uniform rotation collapse simulation.}
\label{fig:Saijo_collapse}
\end{centering}
\end{figure}

Our NR simulations use units where $G = c = 1$, so both spatial distances (e.g. coordinate radius $r$) and time $t$ are expressed in terms of a characteristic mass scale $M$.
In general, $M$ can be correspond to any mass scale we desire; here we choose $M = \MSun$ for definiteness, and this determines the value of the polytropic constant $K$ in code units.
Our 3D computational domain is covered with a Carpet mesh with 7 initial levels of refinement;
the finest of these levels covers a cubical region $\{\pm 23.60 M\}^3$, with a grid spacing of $0.40 M$. The evolution scheme used is a 4th-order Runge-Kutta scheme, with a local timestep $dt$  constrained by the Courant-Friedrichs-Lewy condition to no more than 0.5 times the local gridspacing. As the evolution proceeds, we use the minimum value of the lapse function $\alpha$ as a criterion for additional refinement, with additional levels added at trigger values of 0.6, 0.4, and 0.2. The full set of grid information can be found in Table~\ref{tab:ModelI_refinement}.
Figure~\ref{fig:Saijo_collapse} shows how the baryonic density $\rhob$ and lapse function $\alpha$ evolve  along the $x$-axis during the simulation, from the initial time $t = 0 M$ through to post-collapse times $t > 1040 M$.

\begin{table}
    \centering
    \caption{Grid structure for the basic Saijo Model I evolution, where collapse was induced through a \qty{5}{\percent} reduction in pressure. Each column shows, respectively, the refinement level, the time $t$ at which it was activated, and the grid's extent and spacing in units of $M$. Levels 7--9 were activated sequentially as the minimum lapse dropped below 0.6, 0.4, and 0.2, respectively, serving as an indicator that higher resolution was required.}
    \label{tab:ModelI_refinement}
    \begin{tabular}{CCCC}
      \toprule
      \textbf{Ref.~Lev.}   & $\bm{t_{\rm activation}\ (M)}$ & $\bm{x_{\rm max}\ (M)}$ & $\bm{\Delta x\ (M)}$ \\
      \midrule
        0 & 0.0 & 1228.80 & 25.60\\
        1 &  0.0 & 755.20 &  12.80\\ %640.0
        2 &  0.0 & 377.60 &   6.40\\ %320.0
        3 &  0.0 & 188.80 &   3.20\\ %160.0
        4 &  0.0 &  94.40 &   1.60\\ %80.0
        5 &  0.0 &  47.20 &   0.80\\ % 40.0
        6 &  0.0 &  23.60 &   0.40\\ % 20.0
        7 & 998.0 &  11.80 &   0.20\\ % 10.0
        8 & 1021.4 &   5.90 &   0.10\\ % 5.0
        9 & 1032.4 &   2.45 &  0.05\\ % 2.0
      \bottomrule
    \end{tabular}
\end{table}

\begin{figure}
\begin{centering}
\includegraphics[width=0.8\textwidth]{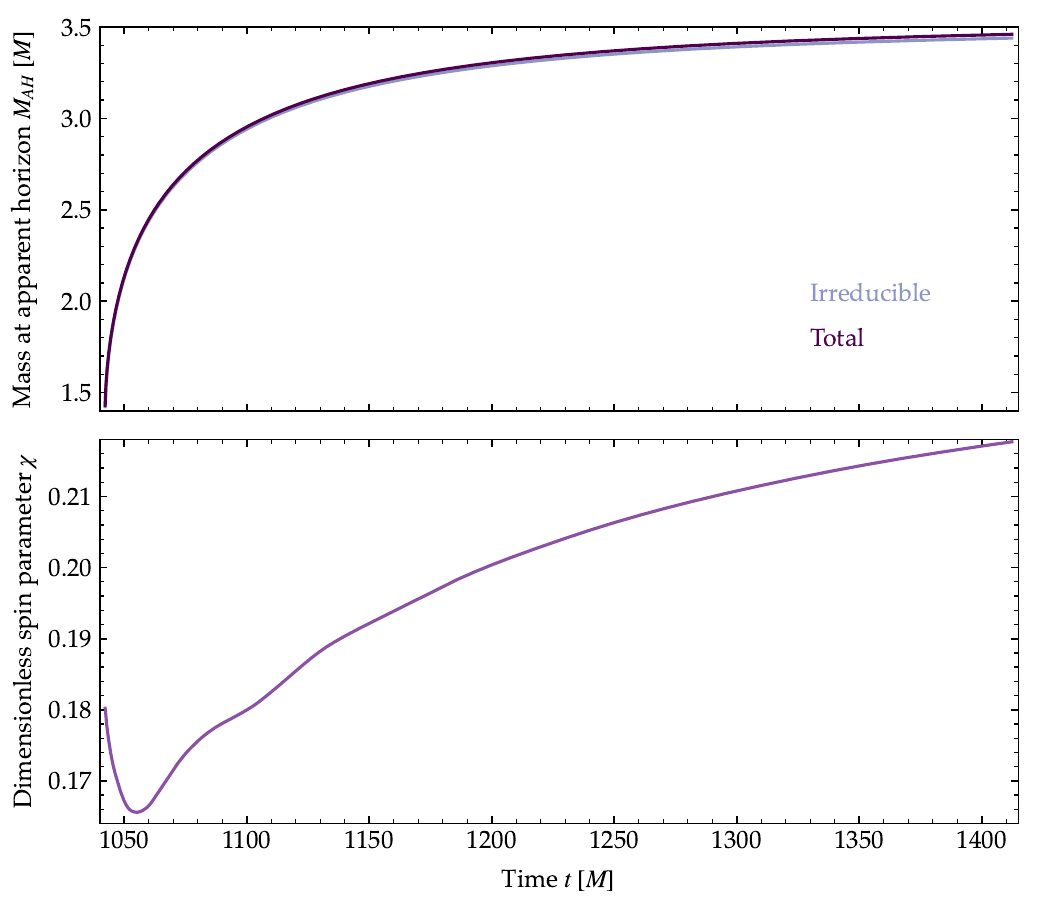}
\caption{\textit{Top}: irreducible and total horizon masses of the DCBH after formation. \textit{Bottom}: dimensionless spin $\chi \equiv J/M_{\rm tot}^2$ after formation.}
\label{fig:Saijo_AHevolve}
\end{centering}
\end{figure}

As the collapse proceeds, curvature in the central region increases until an apparent horizon (AH) is found at $t \approx 1040 M$.
The local grid resolution around the AH is that of refinement level 9 in Table~\ref{tab:ModelI_refinement}, $0.05M$.
The AH is most poorly resolved on the grid at this early time: with an initial minimum radius of $0.657 M$, it is covered by about 24 grid points in each direction.
The AH grows rapidly in coordinate size with increased matter infall and gauge relaxation of the radial coordinate, so that it is always better resolved even after it grows past the boundary of the highest-refined level.
Figure~\ref{fig:Saijo_AHevolve} shows how the total horizon mass $M_{\rm tot}$, the irreducible mass $M_{\rm irr}$, and the spin $\chi \equiv J/M_{\rm tot}^2$ (deduced from the ratio of proper circumference around the equator to that through the poles \cite{Smarr:1973zz}) continue to evolve and grow as more material falls in.
The irreducible mass $M_{\rm irr}$ is defined through the proper area of the horizon: $A = 16 \pi M_{\rm irr}^2$, and it can be related to $M_{\rm tot}$ through the formula \cite{Christodoulou:1970wf}
\begin{eqnarray}
M_{\rm tot}^2 &= \frac{A}{16\pi} + \frac{4 \pi J^2}{A} = M_{\rm irr}^2 + \frac{M_{\rm tot}^4 \chi^2}{4M_{\rm irr}^2} \nonumber \\
\Rightarrow M_{\rm tot}^2 &= \frac{2 M_{\rm irr}^2}{\chi^2} \left( 1 - \sqrt{1 - \chi^2} \right) \rightarrow M_{\rm irr}^2 \left( 1 + \frac{\chi^2}{4} + O(\chi^4) \right), \;\; \chi \ll 1 \nonumber 
\end{eqnarray}
As the spin $\chi$ is low for this configuration, the irreducible and total masses remain close for the duration of the simulation.

% trim = {LEFT, BOTTOM, RIGHT, TOP}

\begin{figure}
\begin{centering}
\includegraphics[width=0.8\textwidth]{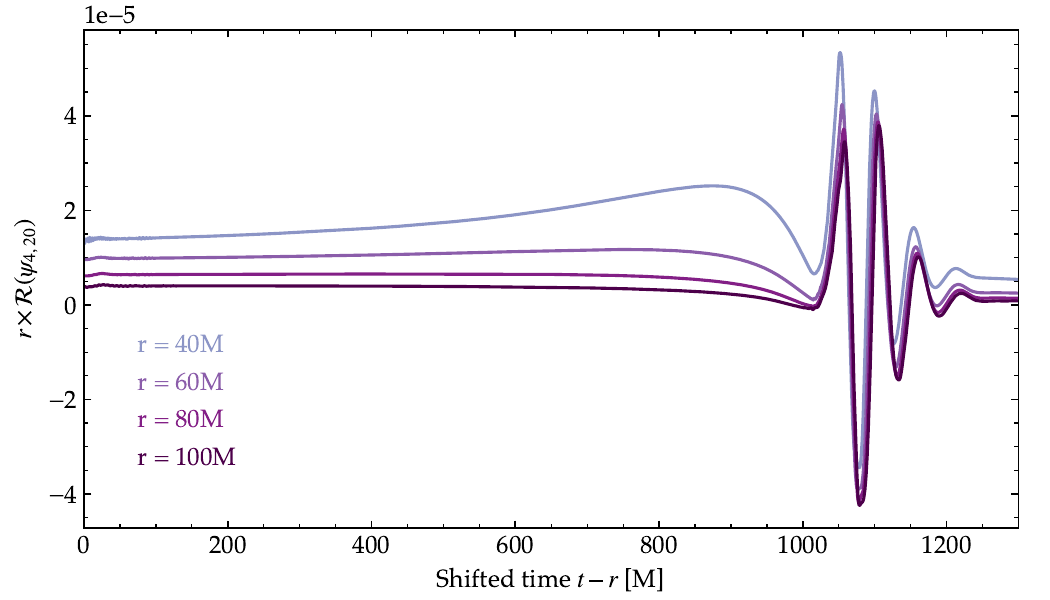}
\caption{Dominant $(l = 2, m = 0)$ mode components of the outgoing Weyl scalar $\psi_4$, extracted at coordinate radii $40 M$, $60 M$, $80 M$, and $100 M$, and translated in time and scaled in amplitude to correct for the different extraction radii. }
\label{fig:Saijo_WFs}
\end{centering}
\end{figure}

Finally, we look at the gravitational waveforms generated from the collapse. These are calculated from the Weyl curvature scalar $\psi_4$, decomposed into spin-weighted spherical harmonics ($s = -2$), and extracted at coordinate radii $40 M$, $60 M$, $80 M$, and $100 M$. Due to the grid refinement structure, the $40 M$ extraction was entirely in a region of grid spacing $0.40 M$, while the $60 M$, $80 M$ extractions had a lower resolution $0.80 M$, and the outermost sphere was only at $1.60 M$.
Due to the axisymmetry, only $m=0$ modes are present in extracted waveforms, and only the leading $(l = 2, m = 0)$ mode shows any significant amplitude. In Figure~\ref{fig:Saijo_WFs} we overlay these waveforms, translating in time to compensate for the travel time to cross each extraction sphere, and scaled by $r$ to compensate for the dominant falloff rate.

For production-quality waveforms, we would need to extrapolate these to infinite radius. However, we are primarily concerned with the qualitative behavior of the simulation at this stage.
We note that waveforms extracted from numerical simulations of compact-object binaries often contain so-called ``junk radiation'' in their early stages.
This typically arises from a combination of factors, including: unresolved constraint violations in the initial data; strong initial localized spacetime curvature due to high momenta and spins of individual compact objects (especially black holes); and a mismatch between ``wave-free'' initial data and the expected history of prior emission from such a configuration.
None of these conditions appears to hold in our current configurations, but we cannot rule out that some junk radiation will emerge as we extrapolate current waveforms and increase resolution. If present, we expect such radiation will occur early in the simulation, and excising the early waveform will be sufficient to remove it.

Note that the final spin of the black hole is modest ($\chi \sim 0.2$), due to our use of uniform rotation, rather than the differential rotation configuration selected by \cite{Saijo:2009ub}. This means that the waveforms produced cannot be meaningfully compared with the results in \cite{Saijo:2009ub}.
Note that the cosmological data profiles of Section~\ref{ssec:cosmo_overview} indicate a differentially rotating fluid, and we have no reason to expect that this will change after the data has been further evolved at the Mesoscale (Section~\ref{ssec:meso_overview}).
However, the current work is intended to show a proof of concept of the full project, and it was more important to demonstrate that we could evolve collapsing rotating data of some type to the point of horizon formation.
The uniform-rotation version of the Saijo data contains less angular momentum overall, and it leads to a less highly spinning black hole after collapse, with a larger horizon and less distortion in the coordinates used for the numerical evolutions. This means that the simulation overall should require less resolution to successfully produce a black hole.
Nevertheless, we have established our ability to perform collapse simulations with Saijo-like data.

The simulations described here were carried out on the machine Pleiades at the NASA Advanced Supercomputing (NAS) facility. Evolution of a configuration to an end time of $1340 M$ took less than 100 hours, using 4 Ivy Bridge nodes, with 4 MPI (Message Passing Interface) processes each. 

\subsection{Sensitivity to Perturbation}

\begin{figure}
\begin{centering}
\includegraphics[width=0.8\textwidth]{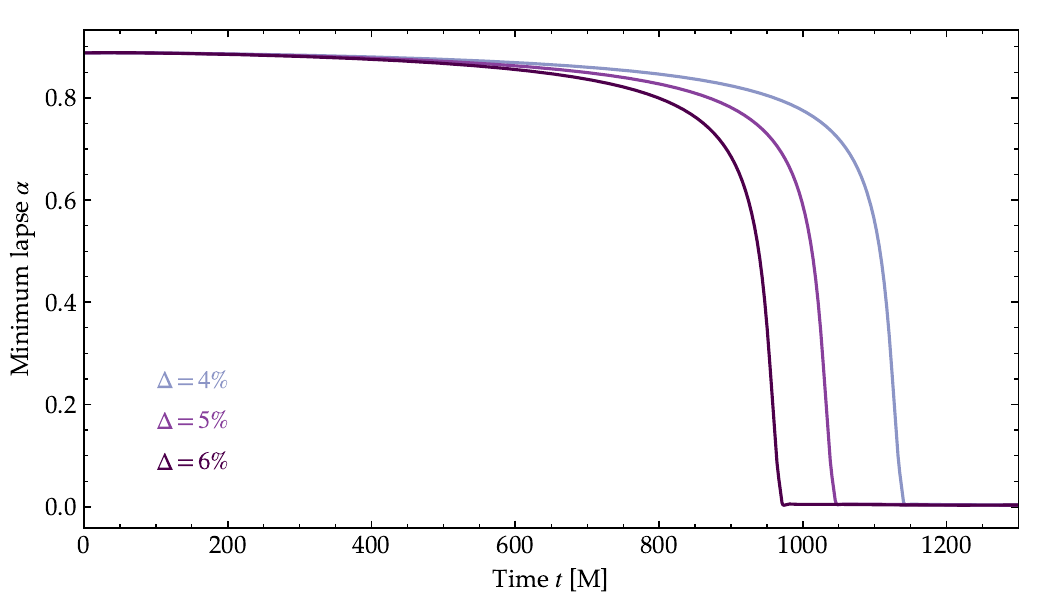}
\caption{Minimum of the lapse function $\alpha$ as a function of evolution time $t$, for three different size perturbations of polytropic $K$.}
\label{fig:Saijo_minlapse_pertsize}
\end{centering}
\end{figure}

\begin{figure}
\begin{centering}
\includegraphics[width=0.8\textwidth]{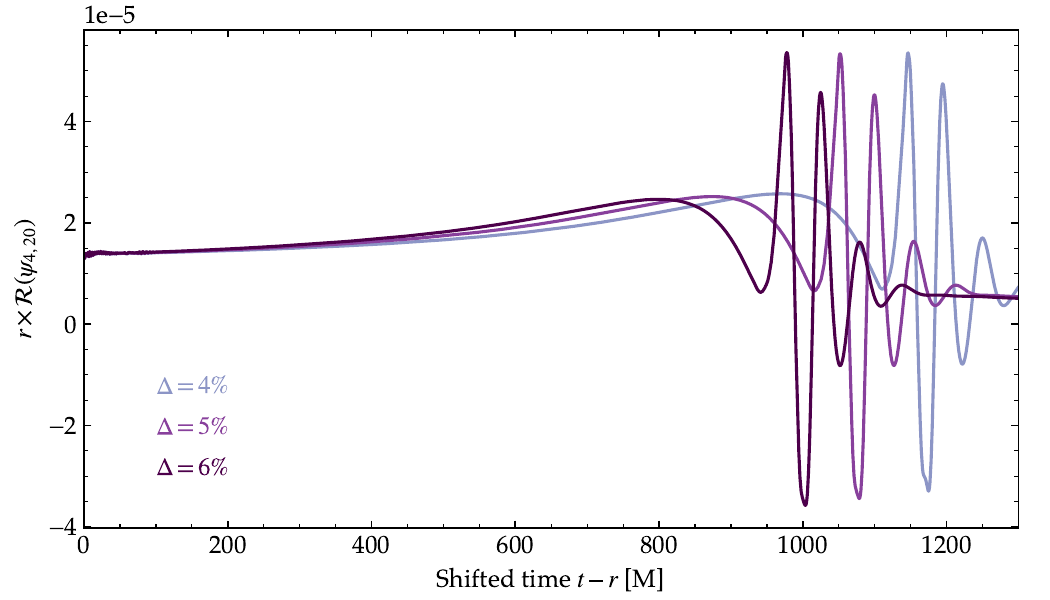}
\caption{Dominant $(l = 2, m = 0)$ mode components of $\psi_4$, extracted at coordinate radius $40 M$, for three different size perturbations of polytropic $K$.}
\label{fig:Saijo_Psi4_20_pertsize}
\end{centering}
\end{figure}

As the method of perturbation used to induce collapse is somewhat arbitrary, we investigate briefly the effect of varying the size of the perturbation of the polytropic $K$.
Figure~\ref{fig:Saijo_minlapse_pertsize} shows how the perturbation size affects the rate of collapse, with more strongly perturbed pressure leading to earlier collapse time.
Nevertheless, the final black-hole formation waveform amplitude is unaffected to leading order, except through its onset time (Fig.~\ref{fig:Saijo_Psi4_20_pertsize}) . 

\section{Summary/Current Status}

Direct-collapse black holes (DCBHs) are relevant to LISA science, both as part of the history of SMBH population growth, and as possible GW sources for LISA in their own formation. In this paper, we have presented early steps in a project to study DCBH formation and the associated GW signals, motivated by
cosmological simulations, and linking multiple computational approaches at different scales.

Many open questions remain as we proceed in this project:

\begin{itemize}
    \item Each computational regime involves considerable complexity not present in the others. Can we capture all the relevant physics when moving from scale to scale (code to code)? Is it necessary to use the full \mesa equation of state in the subsequent NR evolution, or will a simpler approximation (e.g. polytrope or piecewise polytrope) be sufficient, given the shorter timescale of the final collapse?
    \item The requirement of a barotropic (polytropic) equation of state in \rnsid is inconsistent with the shellular rotation assumption of \mesa (i.e. $P = P(\Omega)$). How important is this in practice? Is there a better way to match data from the different regimes? 
    \item How can we best trigger collapse? The \rnsid code assumes a stationarity that will not be present in the \mesa data. Is there a best-possible way to capture the already infalling matter at the start of the NR runs?
    \item  How sensitive will our results be to our choice of each transition time? In principle we can control for this by making the transition from the mesoscale (\mesa code) to the strong-gravity scale (\grhayl) at slightly different times, to determine the most robust time window and the scale of the resulting error in final BH parameters and GWs.
\end{itemize}

If the \rnsid ansatz proves too restrictive, we may consider moving to a more direct use of \mesa profile data, and insert an angular dependence guided by known analytic “thick-disk” stationary configurations (e.g. the “Polish doughnut” of \cite{Paczynski:1979rz}). Alternatively, we can employ a coordinate  transformation like that used to convert the Kerr black hole metric from Boyer-Lindquist form (where the singularity $r = 0$ appears pointlike and the horizon is a coordinate sphere) to Kerr-Schild form (where the singularity appears as a ring in the equatorial plane, and the horizon is an oblate spheroid).

We've concentrated here on issues encountered in initializing and evolving supermassive star configurations to black-hole formation. However, much also remains to be done in obtaining and interpreting cosmological simulation data from \rensims to seed mesoscale evolutions, and to perform data analysis assessments on the gravitational waveforms produced from the final collapse in the context of LISA. These will be the subject of future reports.

\ack
The material is based upon work supported by NASA LISA Preparatory Science Grant 80NSSC24K0360, and taking place under the CRESST II Cooperative Agreement, award number 80GSFC24M0006.
JW acknowledges funding support from NSF grant AST-2510197. ZBE also gratefully acknowledges support from NSF awards PHY-2508377, PHY-2409654, OAC-2411068, and AST-2108072/2227080, as well as NASA awards TCAN-80NSSC18K1488, ATP-80NSSC22K1898, and EPSCoR-80NSSC24M0106.
TA is thankful for support from NSF grants OAC-2229652 and AST-2108269.
The NR simulations described here were carried out on the machine Pleiades at the NASA Advanced Supercomputing (NAS) facility.\footnote{https://www.nas.nasa.gov/hecc/resources/pleiades.html}

% \section*{References}

%\bibliographystyle{iopart-num-mod}
%\bibliography{DCBH_references}
\renewcommand{\bibname}{References}
\printbibliography

\end{document}